\documentclass[10pt,aps,prl,twocolumn,superscriptaddress,longbibliography,showpacs]{revtex4-1}
\usepackage{epsfig}
\usepackage{dcolumn}
\usepackage{bm}
\usepackage{amsmath}
\usepackage{amsfonts}
\usepackage{latexsym}
\usepackage{amssymb}
\usepackage{color}
\usepackage{hyperref}
\usepackage[normalem]{ulem}
\usepackage{fancyhdr}
\usepackage{amsbsy}
\usepackage{amsmath}
\usepackage{subfigure}
\usepackage{epsfig}
\usepackage{amssymb}
\usepackage{pifont}
\usepackage{textcomp}
\usepackage{gensymb}
\usepackage{pifont}
\usepackage{enumerate}
\usepackage{color}
\usepackage{graphicx}

\usepackage{tikz,xcolor,hyperref}
\definecolor{lime}{HTML}{A6CE39}
\DeclareRobustCommand{\orcidicon}{%
	\begin{tikzpicture}
	\draw[lime, fill=lime] (0,0)
	circle [radius=0.16]
	node[white] {{\fontfamily{qag}\selectfont \tiny ID}};
	\draw[white, fill=white] (-0.0625,0.095)
	circle [radius=0.007];
	\end{tikzpicture}
	\hspace{-2mm}
}

\foreach \x in {A, ..., Z}{%
	\expandafter\xdef\csname orcid\x\endcsname{\noexpand\href{https://orcid.org/\csname orcidauthor\x\endcsname}{\noexpand\orcidicon}}
}

\newcommand{\orcidauthorA}{0000-0002-5008-8165} 
\newcommand{\orcidauthorB}{0000-0002-1550-5429} 
\newcommand{\orcidauthorE}{0000-0002-6567-6728} 
\newcommand{\orcidauthorG}{0000-0003-2727-1102} 
\newcommand{\orcidauthorH}{0000-0002-4326-0414} 

\begin{document}

\title{Tuning interchain ferromagnetic instability in A$_{2}$Cr$_{3}$As$_{3}$ ternary arsenides \\ by chemical pressure and uniaxial strain}

\author{Giuseppe Cuono\orcidB}
\affiliation{International Research Centre Magtop, Institute of Physics, Polish Academy of Sciences,
Aleja Lotnik\'ow 32/46, PL-02668 Warsaw, Poland}

\author{Filomena Forte}
\affiliation{Consiglio Nazionale delle Ricerche CNR-SPIN, UOS Salerno, I-84084 Fisciano (Salerno),
	Italy}
\affiliation{Dipartimento di Fisica "E.R. Caianiello", Universit\`a degli Studi di Salerno, I-84084 Fisciano
(SA), Italy}

\author{Alfonso Romano\orcidH}
\affiliation{Dipartimento di Fisica "E.R. Caianiello", Universit\`a degli Studi di Salerno, I-84084 Fisciano
(SA), Italy}
\affiliation{Consiglio Nazionale delle Ricerche CNR-SPIN, UOS Salerno, I-84084 Fisciano (Salerno),
Italy}

\author{Xing Ming\orcidE}
\affiliation{College of Science, Guilin University of Technology, Guilin 541004, PR China}

\author{Jianlin Luo}
\affiliation{Beijing National Laboratory for Condensed Matter Physics and Institute of Physics, Chinese Academy of Sciences,
Beijing 100190, China}
\affiliation{Songshan Lake Materials Laboratory, Dongguan, Guangdong 523808, China}
\affiliation{School of Physical Sciences, University of Chinese Academy of Sciences, Beijing 100190, China}

\author{Carmine Autieri\orcidA}
\email{autieri@magtop.ifpan.edu.pl}
\affiliation{International Research Centre Magtop, Institute of Physics, Polish Academy of Sciences,
	Aleja Lotnik\'ow 32/46, PL-02668 Warsaw, Poland}
\affiliation{Consiglio Nazionale delle Ricerche CNR-SPIN, UOS Salerno, I-84084 Fisciano (Salerno),
	Italy}

\author{Canio Noce\orcidG}
\affiliation{Dipartimento di Fisica "E.R. Caianiello", Universit\`a degli Studi di Salerno, I-84084 Fisciano
(SA), Italy}
\affiliation{Consiglio Nazionale delle Ricerche CNR-SPIN, UOS Salerno, I-84084 Fisciano (Salerno),
Italy}

\pacs{71.15.-m, 71.15.Mb, 75.50.Cc, 74.40.Kb, 74.62.Fj, 62.20.-x}

\date{\today}
\begin{abstract}
We analyze the effects of chemical pressure induced by alkali metal substitution and uniaxial strain on magnetism in the A$_2$Cr$_3$As$_3$ (A = Na, K, Rb, Cs) family of ternary arsenides with quasi-one dimensional structure. Within the framework of the density functional theory, we predict that the non-magnetic phase is very close to a 3D collinear ferrimagnetic state, which realizes in the regime of moderate correlations, such tendency being common to all the members of the family with very small variations due to the different interchain ferromagnetic coupling. 
We uncover that the stability of such interchain ferromagnetic coupling has a non-monotonic behavior with increasing the cation size, being critically related to the degree of structural distortions which is parametrized by the Cr-As-Cr bonding angles along the chain direction. In particular, we demonstrate that it is boosted in the case of the Rb, in agreement with recent experiments.
We also show that uniaxial strain is a viable tool to tune the non-magnetic phase towards an interchain ferromagnetic instability. The modification of the shape of the Cr triangles within the unit cell favors the formation of a net magnetization within the chain and of a ferromagnetic coupling among the chains. This study can provide relevant insights about the interplay between superconductivity and magnetism in this class of materials.
\end{abstract}

\maketitle

\section{Introduction}

A great attention has been devoted in the last years to the competing quantum orders which develop in transition metal pnictides. In this context, one of the most relevant examples is represented by the recently discovered Cr-based arsenide family A$_{2}$Cr$_{3}$As$_{3}$  (A = Na, K, Rb, and Cs) \cite{Mu18,Bao15,Tang15,Tang15b}. This series has been shown to represent an ideal platform for the study of the interplay between magnetism and structural instabilities \cite{Taddei17,Taddei18,Noce20,Wu15,Wu19}, as well as of other relevant features such as noncentrosymmetry and quasi-one-dimensional properties \cite{Kong15,Watson17}.

\begin{figure*}[]
\centering
\includegraphics[scale=0.45, angle=0]{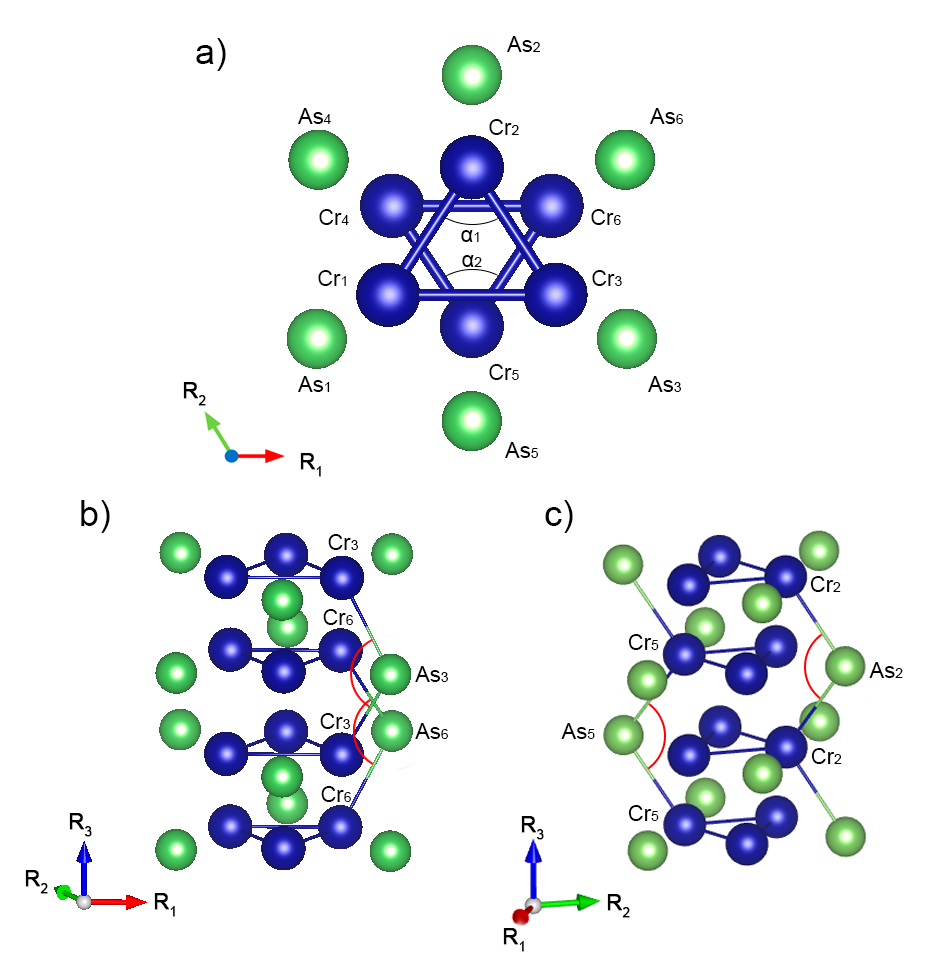}
\caption{(a) Cr-triangles belonging to the {[}(Cr$_{3}$As$_{3}$)$^{2-}${]}$_{\infty}$ sub-nanotubes of the A$_{2}$Cr$_{3}$As$_{3}$ compounds. The deviation from the ideal hexagonal structure is parametrized by the angles $\alpha_1$ and $\alpha_2$. Blue and green circles denote Cr and As atoms, respectively. (b-c) Cr-As-Cr bonding angles along different chain direction (see the reported axes).
}
\label{Triangle}
\end{figure*}

The crystal structure of the above-mentioned compounds, reported in Fig.~\ref{Triangle}, features Cr$_3$As$_3$ chains forming double-wall sub-nanotubes (DWSN), where the inner-wall tubes are constructed by Cr triangles and the outer-wall tubes by As triangles, separated by columns of A$^{+}$ ions \cite{Mu18,Bao15,Tang15,Tang15b}. Band structure calculations indicate that the states governing the electronic physics at the Fermi energy $E_F$ are mainly Cr 3$d$ orbitals \cite{Jiang15,Cuono18,Cuono19c}.

Interestingly, in these compounds a superconducting phase \cite{Zhi15,Adroja15,Pang15} also emerges at ambient pressure, which is so far unique for systems containing chromium  \cite{Noce20,Wu10,Wu14,Kotegawa14,Autieri17,Autieri17b,Autieri18,Cuono19,Cuono19b,Nigro19}. Notably, superconductivity occurs with a background
of ferromagnetic (FM) spin fluctuations \cite{Luo19}, which is quite rare and make a major difference with respect to cuprates, iron-based and heavy-fermion systems where unconventional superconductivity occurs close to an antiferromagnetic (AFM) instability \cite{Goll06}.

Recent NMR measurements reveal the evidence of strong spin fluctuations. In the Rb-based compound, strong enhancement of the spin susceptibility suggests that 3D FM fluctuations are present near $T_c$ \cite{Yang15}. Analogous experiments in K$_{2}$Cr$_{3}$As$_3$ show that Cr spin fluctuations may have a dominant antiferromagnetic character \cite{Zhi15}, while in Na$_{2}$Cr$_{3}$As$_3$  and Cs$_{2}$Cr$_{3}$As$_3$ they are suppressed \cite{Zhi16}. Therefore, a systematic comparison of how alkali substitution modifies the FM fluctuations among the members of the family is needed in order to shed light on the magnetic ground states of these compounds, and possibly on the superconducting pairing mechanism. The latter still remains elusive, with conflicting hypotheses that have been suggested, ranging from spin fluctuations to conventional phonons \cite{Zhi15,Adroja15,Pang15,Taddei17,Subedi15,Zhong15,Wu15b}.

In this paper we report a systematic first-principle study of magnetism in the A$_2$Cr$_3$As$_3$ family and show that the tendency towards an interchain FM instability may be tuned via chemical substitution or uniaxial strain.
Isovalent doping can induce chemical pressure effects without introducing carriers. At the same time it alters the interchain hopping, also affecting the coupling between the chains. Moreover, it may cause slight modifications of the crystal structure along the chain, thus altering the intrachain spin correlations. On the other hand, uniaxial strain, e.g.  the modification of the lattice constant along a specific direction, is effective especially in 1D compounds, since it allows to continually modify the distances parallel or orthogonal to the chain direction, and thus can provide a pathway to explore the related phase diagram.

Recently it has been demonstrated that local deformations \cite{Taddei18} from the ideal hexagonal structure have strong impact on the magnetic properties in those compounds \cite{Cuono20}. In particular, in K$_2$Cr$_3$As$_3$ orthorhombic distortions make the Cr triangles in the DWSN no longer equilateral, which is detrimental to magnetic frustration and thus promotes a new collinear ferrimagnetic ground state with a net magnetization emerging along the chain \cite{Cuono20}.
Moreover, the non-magnetic state is predicted to be in close proximity to a weak 3D ferrimagnetic phase.

Here we perform an analogous systematic study on the members of the A$_2$Cr$_3$As$_3$ family (A=Na, K, Rb, Cs), showing that all the compounds are in the proximity of a collinear ferrimagnetic configuration within the chain, regardless of the A species.  
Our results also demonstrate that interchain ferromagnetic exchange is promoted in the regime of moderate on-site electronic correlation, and that the stability of such ferromagnetic coupling has a 
non-monotonic behavior with the size of the cation. 
We single out the structural parameters that govern the formation of the ferromagnetic coupling, focusing in particular on the Cr-As-Cr bond angles originating from the alternate distribution of alkali metal ions along the chain direction. 
Furthermore, we show that uniaxial strain affects the interchain FM instability by lowering the critical value of the local Coulomb repulsion above which the magnetic moments are stabilized.

The paper is organized as follows. In Sec. II we present the computational details of our approach,  Sec. III is devoted to a systematic study of the magnetic configurations within the chain for all the compounds. In Sec. IV we analyze the effects of the interchain magnetic coupling on the stability of the FM configurations, together with the corresponding structural fingerprints. In Sec. V we consider the case of a compressive/tensile strain field applied along an in-plane direction. Finally, Sec. VI is devoted to the conclusions.

\section{Computational details}

We have performed density function theory (DFT) calculations by using the VASP package \cite{Kresse93,Kresse96,Kresse96b} based on the
plane wave basis set and the Projector Augmented Wave \cite{Kresse99} method with a cutoff of 440 eV for the plane wave basis.
We have used the PBEsol exchange-correlation method \cite{Perdew08}, that provides an excellent functional for the atomic relaxation in solids.
We need a good accuracy for the structural properties, given the strong coupling  between electronic and structural degrees of freedom in this class of materials.
A 4$\times$4$\times$10 $k$-point grid, which corresponds to 160 $k$-points in the first Brillouin zone, has been employed for the calculations concerning the chains.
In order to describe the electronic correlations associated with the Cr 3$d$ states, a Coulomb repulsion $U$ has been added to our functional  \cite{Liechtenstein95}.
We have used values of $U$ ranging from 0 to 3 eV.
The relevant region is until 1.5 eV, that is, for larger values of $U$ we are in the saturation region.
We have performed structural relaxation minimizing the internal atomic positions and forces to less than 0.01 eV/{\AA}.
The volume calculated in DFT usually deviates from the experimental value by a few percent. While in most of the cases this deviation is negligible, it could play a relevant role in systems such as the one investigated here, characterized by a strong interplay between magnetic and structural properties. 
Though in other studies on Cr-based compounds a full lattice relaxation procedure has been adopted\cite{Wu15,Jiang15,Cao15}, we decided to perform the relaxation of the atomic positions having fixed the volume of the unit cell to its experimental value. In this way we exclude the possibility of volume variations which would affect the magnetic properties of the system. We have also checked that when the constraint on the volume is removed, our results coincide for instance with those obtained in Ref.    \onlinecite{Cao15}. 
\\

The relaxation procedure started from equilateral triangles for both collinear and non-collinear magnetic configurations, also including the effect of the spin-orbit coupling. We found that once the structural relaxation converges, the energetically favored configuration is characterized by distorted Cr-ion triangles, with an arrangement of the magnetic moments which is always collinear.
The lattice vectors of the unit cell are \textbf{R}$_1$=(a,0,0), \textbf{R}$_2$=(-a/2,$\sqrt{3}a/2$,0), and \textbf{R}$_3$=(0,0,c).
To calculate the interchain coupling we have doubled the unit cell along the \textbf{R}$_1$ direction,  and we have used a supercell with two formula units and a 2$\times$4$\times$10 $k$-point grid.

\section{Chain ground state in the A$_{2}$C\lowercase{r}$_{3}$A\lowercase{s}$_{3}$ family}

The A$_{2}$Cr$_{3}$As$_{3}$ family is characterized by a markedly 1D structure, with Cr$_3$As$_3$ linear chains separated by A$^{+ }$ ions sitting on different crystallographic
sites \cite{Mu18,Bao15,Tang15,Tang15b}. In Fig.~\ref{Triangle} we report the notation used for the inequivalent Cr and As atoms, together with the angles among the Cr atoms within the \textbf{R}$_1$-\textbf{R}$_2$ plane and the Cr-As-Cr bonding angles along the chain. Increasing the atomic radius from Na$^{+ }$ to K$^{+ }$ , Rb$^{+ }$  and Cs$^{+ }$  induces chemical pressure effects which lead to the expansion of the interchain distance by keeping the linear chain structure almost unchanged, as one can see from the data reported in Table \ref{Tab1}.

We note that the superconducting critical temperature $T_c$ progressively increases in the series A$_2$Cr$_3$As$_3$ as the atomic number of the alkaline ion decreases. Actually, $T_c$ is found to be $\sim$ 2.2 K in Cs$_{2}$Cr$_{3}$As$_3$ \cite{Tang15b}, 4.8 K in Rb$_{2}$Cr$_{3}$As$_3$ \cite{Tang15}, 6.1 K in K$_{2}$Cr$_{3}$As$_3$ \cite{Bao15}, and 8.6 K in Na$_{2}$Cr$_{3}$As$_3$ \cite{Mu18}.
This finding thus clearly indicates a systematic trend of the chemical pressure effect on the superconducting mechanism \cite{Bao15,Tang15,Tang15b,Mu18}.
Of course, reducing the A-cation size, the unit-cell volume decreases too. However, it has been shown that the application of hydrostatic pressure on K$_2$Cr$_3$As$_3$, which differently from the chemical pressure reduces the volume isotropically, leads to a reduction of $T_c$ \cite{Sun17,Kong15,Wang16}.
This means that the two kinds of pressure have opposite effects on the superconducting transition.

From the structural point of view the chemical pressure may affect other relevant structural parameters,  which are expected to play a role in the most favorable magnetic stable phases.	We know from previous ab-initio studies on K$_{2}$Cr$_{3}$As$_{3}$ \cite{Taddei18,Cuono20} that localized orthorhombic distortions of the CrAs sublattice together with associated K displacements cause the Cr-triangles in the DWSN to be no longer equilateral. These deformations, described in terms of the angles $\alpha_1$ and $\alpha_2$ reported in Fig.~\ref{Triangle} weaken magnetic frustration, favoring a collinear stripe phase within the chains \cite{Cuono20}. The structural distortions explicitly taken into account in our approach make the two magnetic moments at the basis of each triangle different from the one at the vertex.

\noindent 
\begin{table}[t!]
\begin{centering}
\begin{tabular}{|c|c|c|c|c|}
\hline 
   A$_{2}$Cr$_{3}$As$_{3}$  & A=Na \cite{Mu18}  & A=K \cite{Bao15}  & A=Rb \cite{Tang15}  & A=Cs \cite{Tang15b}   \tabularnewline
\hline 
a(\AA)  & 9.2390  & 9.9832  & 10.2810  & 10.6050  \tabularnewline
\hline 
c(\AA)  & 4.2090  & 4.2304  & 4.2421  & 4.2478  \tabularnewline
\hline  
\end{tabular}
\par\end{centering}
\caption{Lattice constants of the compounds of the family A$_{2}$Cr$_{3}$As$_{3}$. The space group is the P$\bar{6}$m2 (No. 187). }
\label{Tab1} 
\end{table}

However, other parameters are worth paying attention to. The alternate distribution of the alkali metal ions along the chain direction gives rise to different bond angles of Cr-As-Cr, as shown in Fig.~\ref{Triangle}b. Due to the distorted Cr triangles, we can distinguish four independent angles, Cr$_3$-As$_3$-Cr$_3$, Cr$_6$-As$_6$-Cr$_6$, Cr$_2$-As$_2$-Cr$_2$ and Cr$_5$-As$_5$-Cr$_5$ . In the following, we study the magnetic phases and the way they get correlated to structural deformations parametrized by the in-plane angles between the Cr atoms in the isosceles triangles and the different Cr-As-Cr bonding angles. The analysis is performed as a function of the Coulomb repulsion $U$, for different choices of the chemical substitution.

\subsection{Chain ground state of C\lowercase{s}$_{2}$C\lowercase{r}$_{3}$A\lowercase{s}$_{3}$}
\begin{figure*}
  \includegraphics[scale=0.35]{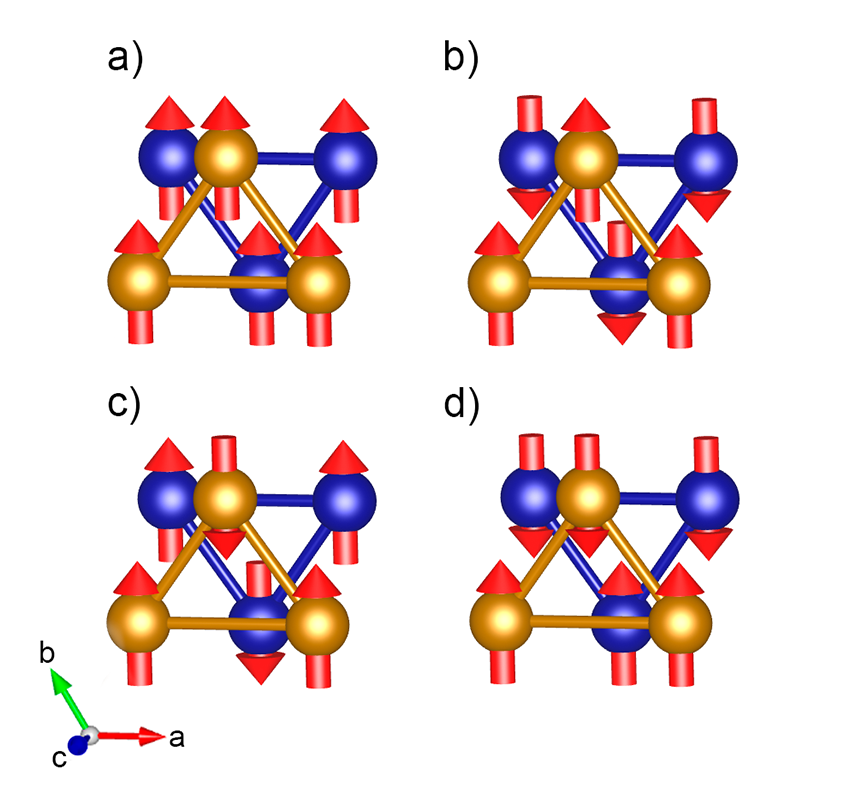}
	\caption{Cr-spin configurations investigated in the paper, defined as: a) the ferromagnetic state (FM), b) the interlayer antiferromagnetic state  (AFM), c) the up-up-down/up-up-down  ($\uparrow\uparrow\downarrow$-$\uparrow\uparrow\downarrow$) stripe state, d) the up-up-down/down-down-up  ($\uparrow\uparrow\downarrow$-$\downarrow\downarrow\uparrow$) zig-zag state. The two different colors, blue and yellow, distinguish the Cr-atoms at the two different planes.
	}
	\label{configurations}
\end{figure*}

We apply the procedure already presented in Ref. \cite{Cuono20} for the K$_{2}$C\lowercase{r}$_{3}$A\lowercase{s}$_{3}$ to the representative case of the C\lowercase{s}$_{2}$C\lowercase{r}$_{3}$A\lowercase{s}$_{3}$, and perform the atomic relaxation at different values of the Coulomb repulsion $U$ for the distorted triangles composed of Cr-atoms and belonging to the DWSN.
We consider four possible collinear magnetic states, which for moderate values of $U$ turn out to be
the magnetic stable states. They are reported in Fig. \ref{configurations} and are classified as the ferromagnetic state (FM), the interlayer antiferromagnetic state (AFM), the up-up-down/up-up-down stripe state ($\uparrow\uparrow\downarrow$-$\uparrow\uparrow\downarrow$) and the up-up-down/down-down-up zig-zag state ($\uparrow\uparrow\downarrow$-$\downarrow\downarrow\uparrow$). Together with these states, we also consider the non-magnetic (NM) one.

\begin{figure}[]
\centering
\includegraphics[width=\columnwidth, angle=0]{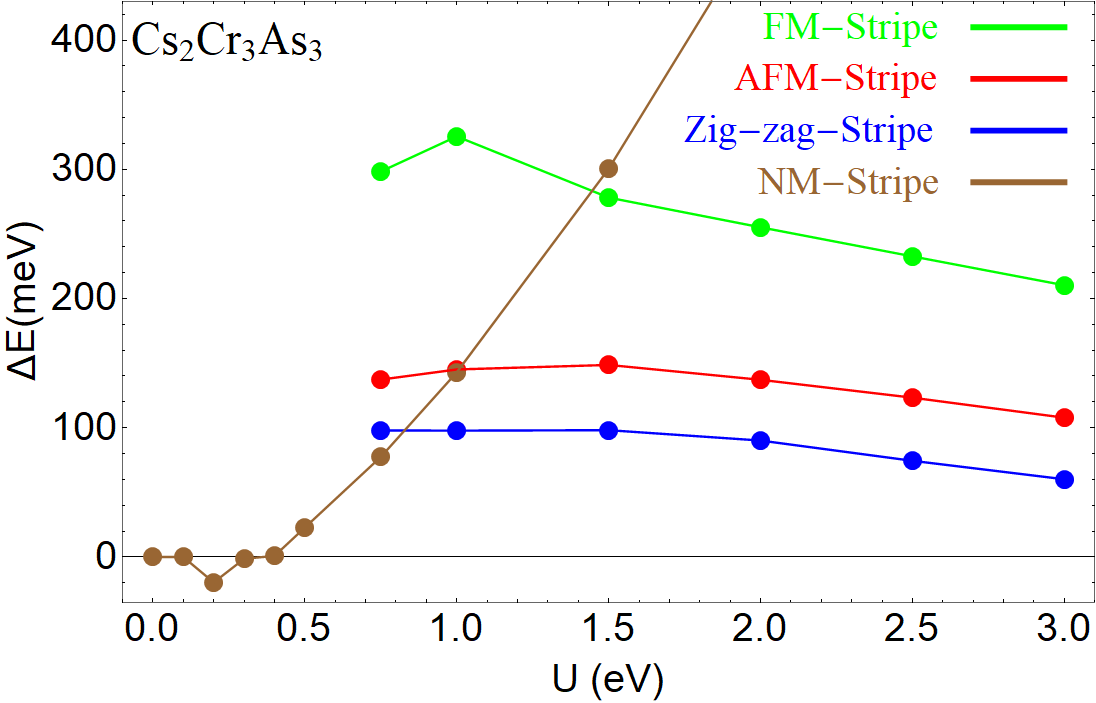}
\caption{Energies per Cr-atom of the FM, AFM, zig-zag and NM states measured with respect to the stripe state energy, plotted as functions of the Coulomb repulsion for the compound Cs$_2$Cr$_3$As$_3$.}
\label{EC}
\end{figure}

In Fig.~\ref{EC} the energy difference between the NM, FM, AFM, and the zig-zag state with respect to the stripe state ($\uparrow\uparrow\downarrow$-$\uparrow\uparrow\downarrow$) displays the existence of two regimes: for values of the Coulomb repulsion $U<U_c=0.4\,$eV, the ground state is non-magnetic, then becoming the stripe one for $U\geq U_c$. In the following, we will also refer to the stripe groundstate as ferrimagnetic, standing for a collinear configuration where the magnetic moments are ucompensated within the unit cell, and the resulting moment is due the majority spins of the Cr ions at the basis of each triangle. \\

\begin{figure}
	\centering
	\includegraphics[width=\columnwidth, angle=0]{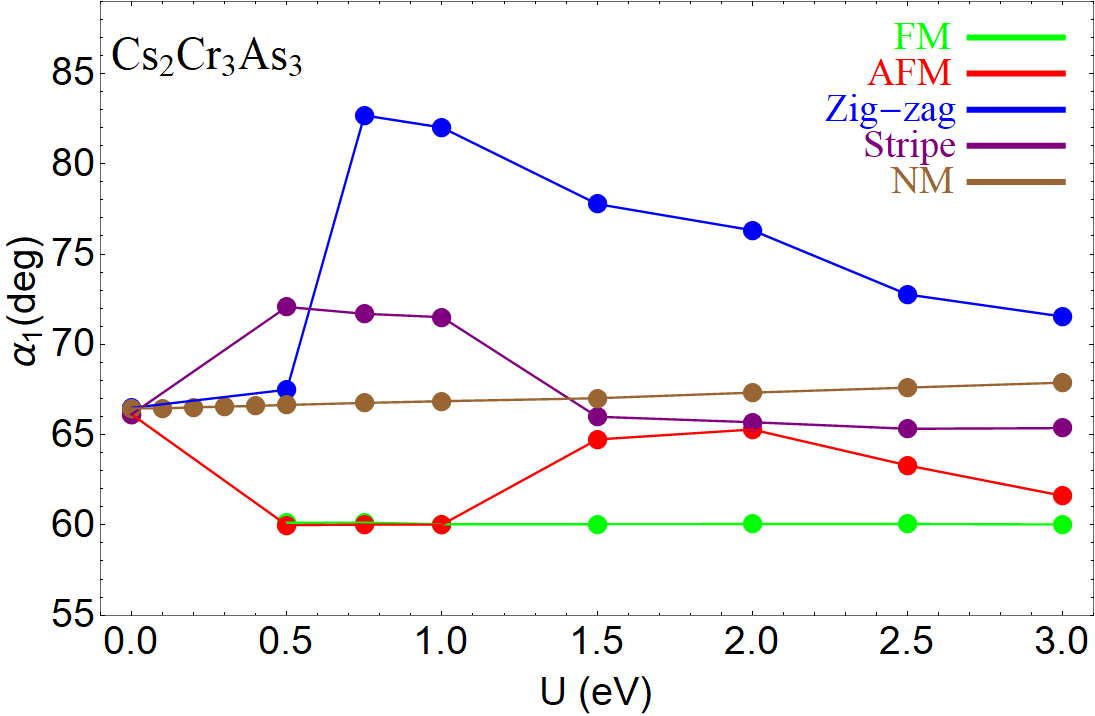}
	\caption{Angle $\alpha_{1}$ as a function of the Coulomb repulsion for the compound Cs$_2$Cr$_3$As$_3$.}
	\label{angles}
\end{figure}

\begin{figure}[]
\centering
\includegraphics[width=\columnwidth,  angle=0]{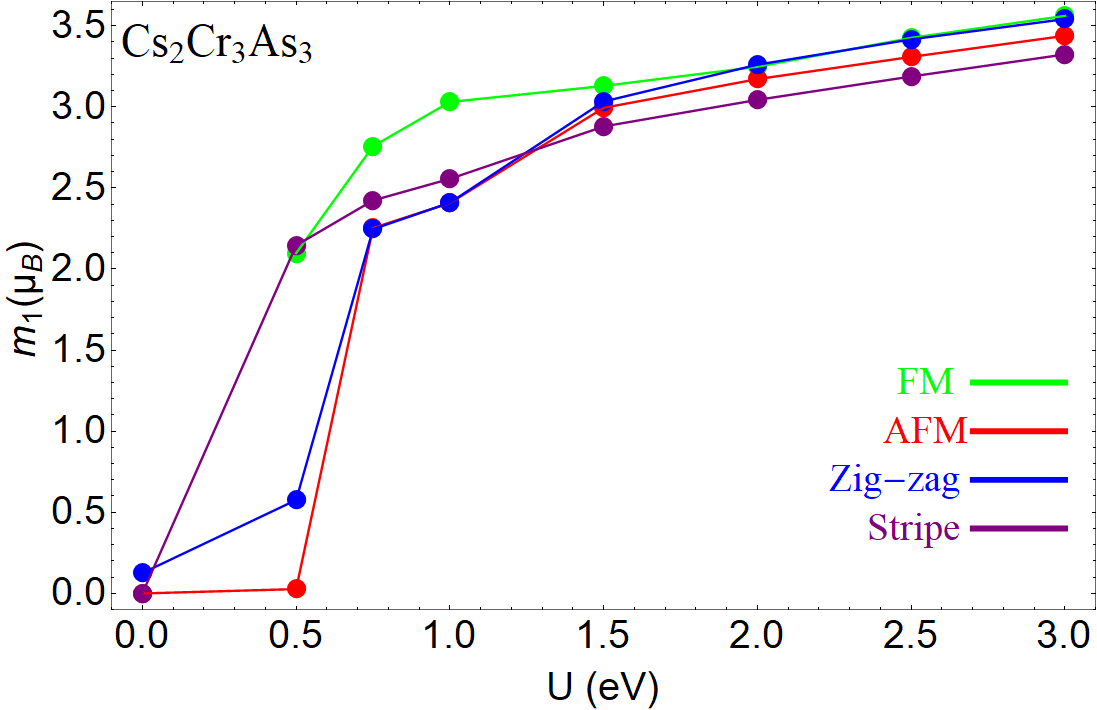}
\caption{Magnetic moment of the Cr$_{1}$ ion as a function of the Coulomb repulsion for the compound Cs$_2$Cr$_3$As$_3$.}
\label{magneticmoment}
\end{figure}

In Figs.~\ref{angles} and \ref{magneticmoment} we report the behavior of the representative angle $\alpha_{1}$ shown in Fig.~\ref{Triangle} and the magnetic moment $m_{1}$ at the Cr$_{1}$ site as functions of $U$. The other angles and magnetic moments have a very similar trend and thus are not reported here. We can see from Fig.~\ref{angles} that $\alpha_{1}$ is always away from the ideal 60$^{\circ}$ value, even in the NM phase, the only exception being represented by the FM state. 
In particular, while in the magnetic phases $\alpha_1$ is characterized by a non-monotonic behavior as $U$ is varied, on the contrary in the NM phase $\alpha_1$ stays almost constant, since there is no interplay between structural and magnetic properties.
In particular we note that in the stripe phase $\alpha_{1}$ shows a sudden jump to $\sim 70^\circ$ at $U \simeq U_c$, and becomes practically constant above $U=1.5\,$eV, differently from what happens in the other magnetic configurations.

The behavior of the magnetic moment at the Cr$_{1}$ site shown in Fig.~\ref{magneticmoment} confirms the strong interplay between structural and spin-orbital degrees of freedom in this class of compounds.
As for the case of the angle $\alpha_1$, a different behavior is found in the two regimes corresponding to values of $U$ approximately lower and higher than 1.5 eV, respectively. At low values of $U$ the magnetic moment tends to vanish in the zig-zag as well as in the stripe phase, thus providing evidence of a non-magnetic ground state configuration.

\subsection{Chain ground state for different cations}
In this subsection we report a systematic comparison of the magnetic stable phases, the $\alpha$ angles and the magnetic moments for all the compounds of the family, as obtained via first-principles analysis in the same range of $U$ and by following the same procedure.

We first point out that all the compounds show a similar behavior, i.e. the existence of a critical $U_c$ separating the NM state ($U<U_c$) from the stripe one ($U>U_c$). We choose to focus on the most relevant comparative outcome and postpone to Appendix I the detailed results.

\begin{figure}[]
\centering
\includegraphics[width=\columnwidth, angle=0]{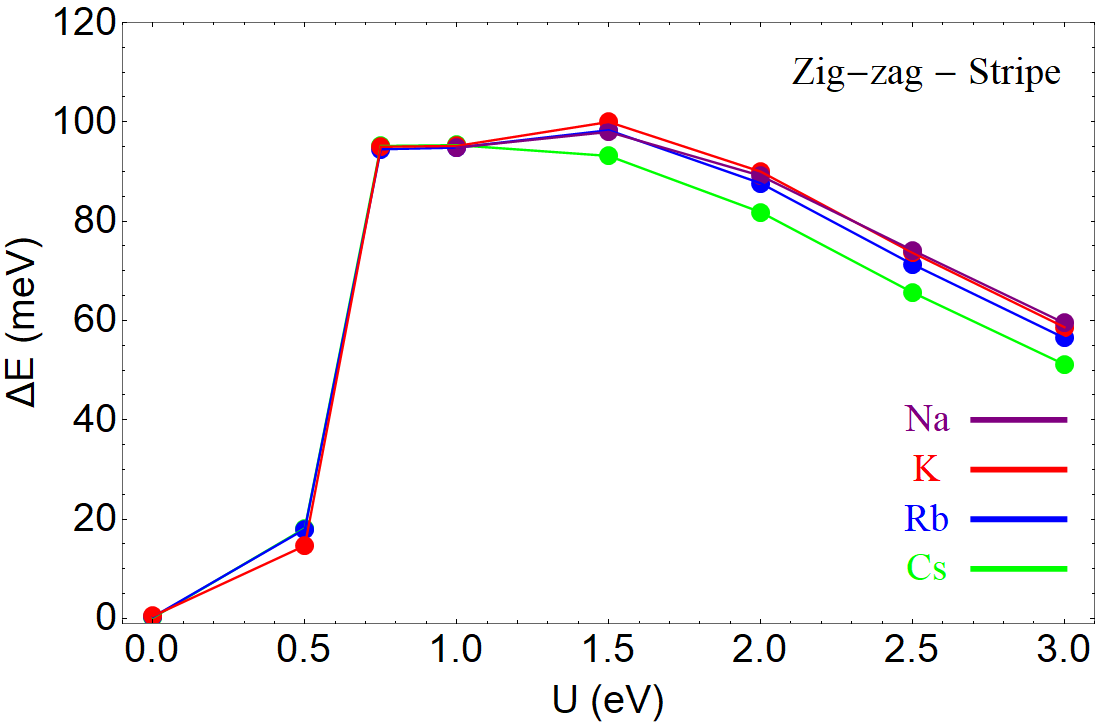}
\caption{Difference between the energies per Cr-atom of the zig-zag and of the stripe state as a function of the Coulomb interaction for the compounds of the family A$_{2}$Cr$_{3}$As$_{3}$ (A=Na, K, Rb, Cs).}
\label{EC_comparison}
\end{figure}

\begin{figure}
	\centering
	\includegraphics[width=\columnwidth, angle=0]{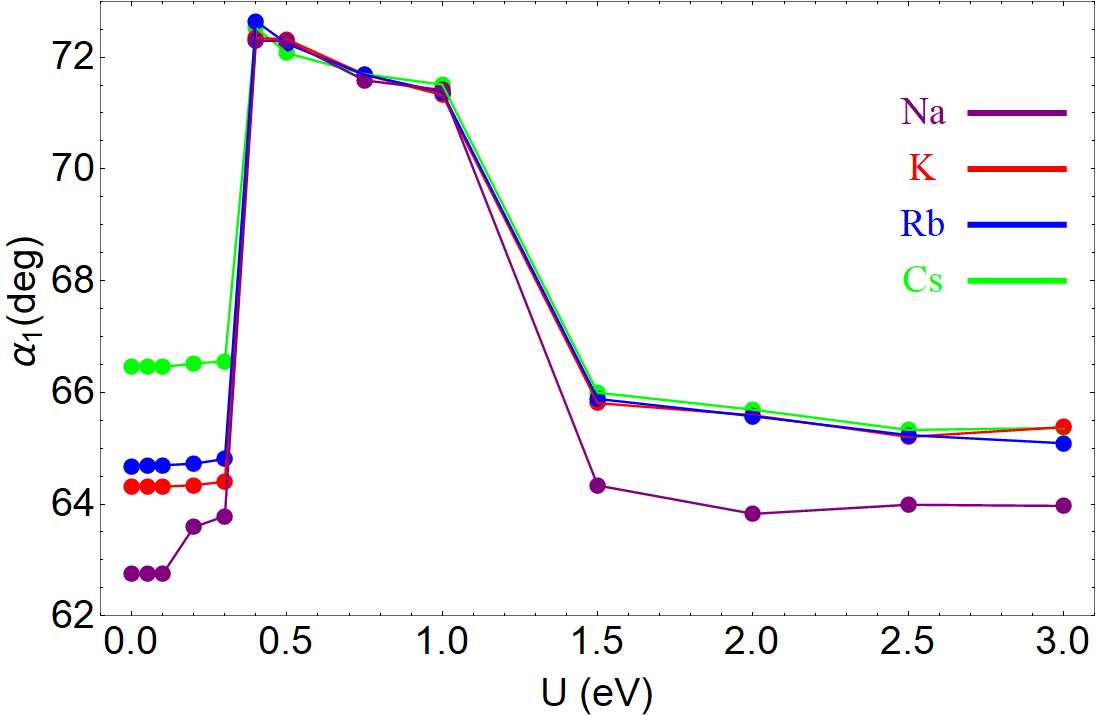}
	\caption{Angle $\alpha_1$ in the ground state as a function of the Coulomb repulsion for the compounds of the family A$_{2}$Cr$_{3}$As$_{3}$ (A=Na, K, Rb, Cs).}
	\label{angles_comparison}
\end{figure}

\begin{figure}[]
\centering
\includegraphics[width=\columnwidth, angle=0]{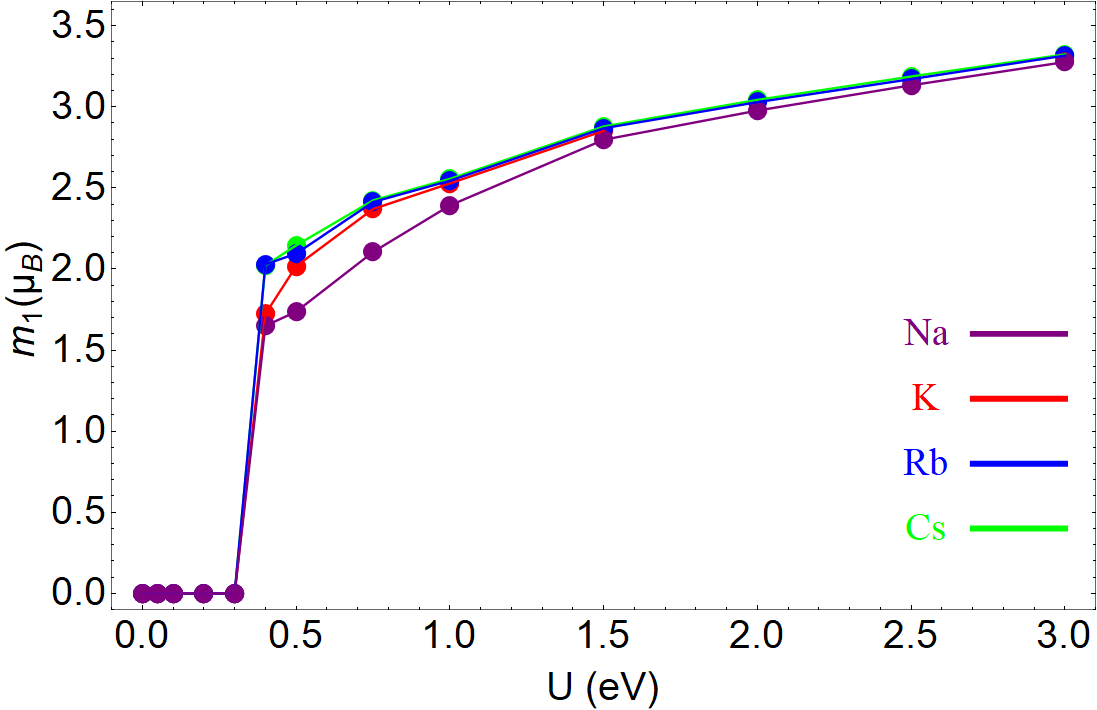}
\caption{Magnetic moment $m_{1}$ in the ground state as a function of
the Coulomb repulsion for the compounds of the family A$_{2}$Cr$_{3}$As$_{3}$ (A=Na, K, Rb, Cs).}
	\label{m_comparison}
\end{figure}

Fig.~\ref{EC_comparison} shows that chemical pressure slightly affects the $U$-dependence of the energy difference between the zig-zag and the ground stripe phase, which only shows a little reduction as the atomic radius increases. In Figs.~\ref{angles_comparison} and \ref{m_comparison} we report the comparison of the angles and the magnetic moments in the ground state for the different atomic species. We see that the Cr triangles are quite distorted in the ground phase, especially at low and intermediate values of $U$, this being a common trend for all the compounds. The distortion is slightly less pronounced only in the case of Na for $U>1.5\,$eV.
The magnetic moment tends to grow with the ion size, saturating
for larger values of $U$ to a value that is very close to the maximum predicted value for Cr ions in the K$_{2}$Cr$_{3}$As$_{3}$ compound, that is $\frac{10}{3}\approx{3.33}$ $\mu_B$. This is so because the oxidation of chromium is +2/3 in K$_{2}$Cr$_{3}$As$_{3}$.

We point out that such collinear ferrimagnetic phase may be qualitatively interpreted \cite{Cuono20} within a minimal effective Heisenberg model which assumes the Cr spins lying in the \textbf{R}$_1$-\textbf{R}$_2$ plane coupled via dominant independent AFM exchange between nearest neighbor Cr atoms and smaller AFM coupling between next nearest neighbor Cr atoms along the \textbf{R}$_3$-axis. Specifically, the onset of the stripe phase is governed by the ratio among the in-plane exchange parameters along the \textbf{R}$_1$ and \textbf{R}$_2$ axes, which in turn is critically linked to the degree of deformation of the triangles, growing with the $\alpha$ angles.

\section{Strength of the magnetism and interchain FM instability in the A$_{2}$C\lowercase{r}$_{3}$A\lowercase{s}$_{3}$ family}
To understand the effect of the chemical pressure due to alkali atom substitution on the most favorable magnetic configuration in these Q1D systems, it is important to consider the
coupling between the chains, which can provide useful insights for the interpretation of the experimental data \cite{Zhi15,Yang15,Zhi16}. In the previous section, we have shown that all the compounds have collinear magnetism above a characteristic value of $U_c$, where the ground state changes from a non-magnetic to a collinear stripe configuration within the DWSN, allowing to attribute a net magnetic moment to each chain. 
Single interchain magnetic interactions then couple spins of neighboring chains. Without interchain magnetic coupling, the system can only show magnetic order in 1D, this being difficult to achieve because of the Mermin-Wagner theorem \cite{Gelfert01,Noce06}. With the presence of the interchain magnetic coupling, the system is not one-dimensional anymore and a magnetic order can more easily develop. This crucial interchain magnetic coupling can be FM or AFM.

We study a supercell of two chains, assuming that in the magnetic phase the coupling within a single chain leads to the stripe configuration and that a macrospin can be associated to each chain.

As a first step, we proceed by looking at the energy difference between the NM, FM and AFM interchain configurations. The study (not reported here for brevity) confirms that up to a critical value of the Coulomb repulsion the NM configuration is the ground state, and that the results for different cations are quite similar. The FM interchain configuration, where the macrospins of the chains are oriented in the same direction, is always lower in energy than the AFM one, confirming that all these compounds in this region are non-magnetic but on the verge of magnetism, sustaining interchain FM spin fluctuations.

\begin{figure}[]
\centering
\includegraphics[width=\columnwidth, angle=0]{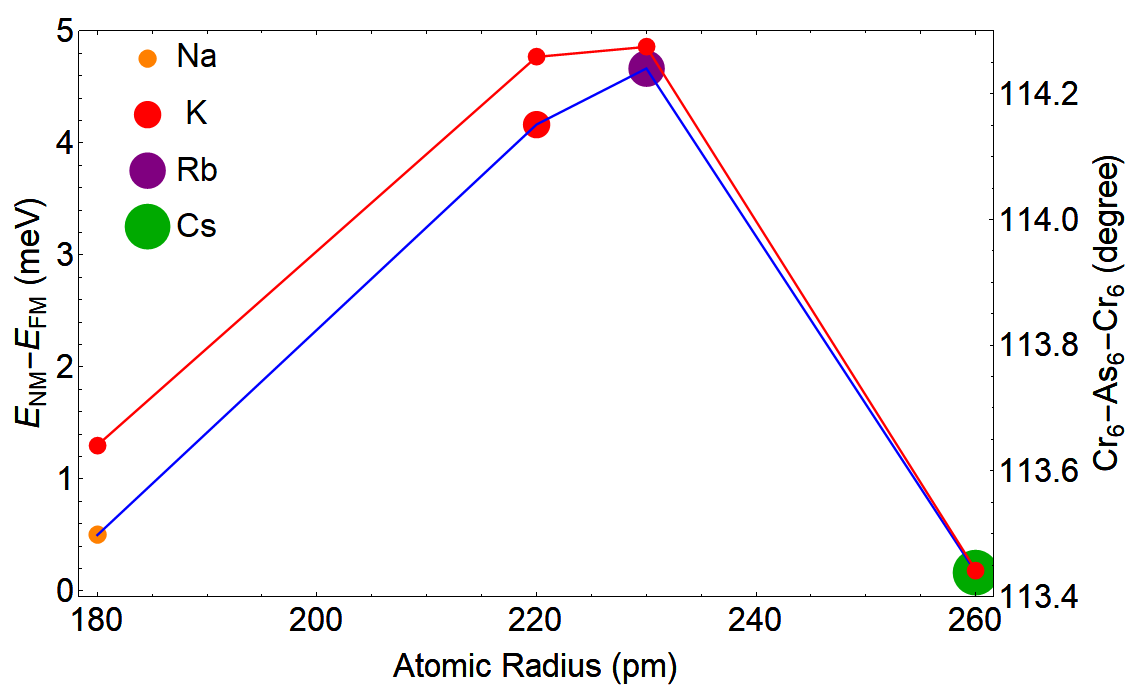}
\caption{Energy difference between the NM and the interchain FM state (blue line, left axis) and bonding angle Cr$_6$-A$_6$-Cr$_6$ (red line, right axis) for different choices of the cation. The value of the Coulomb repulsion is $U=0.75\,$eV.}
\label{EnvsBonding}
\end{figure}

\begin{figure}[]
\centering
\includegraphics[width=\columnwidth, angle=0]{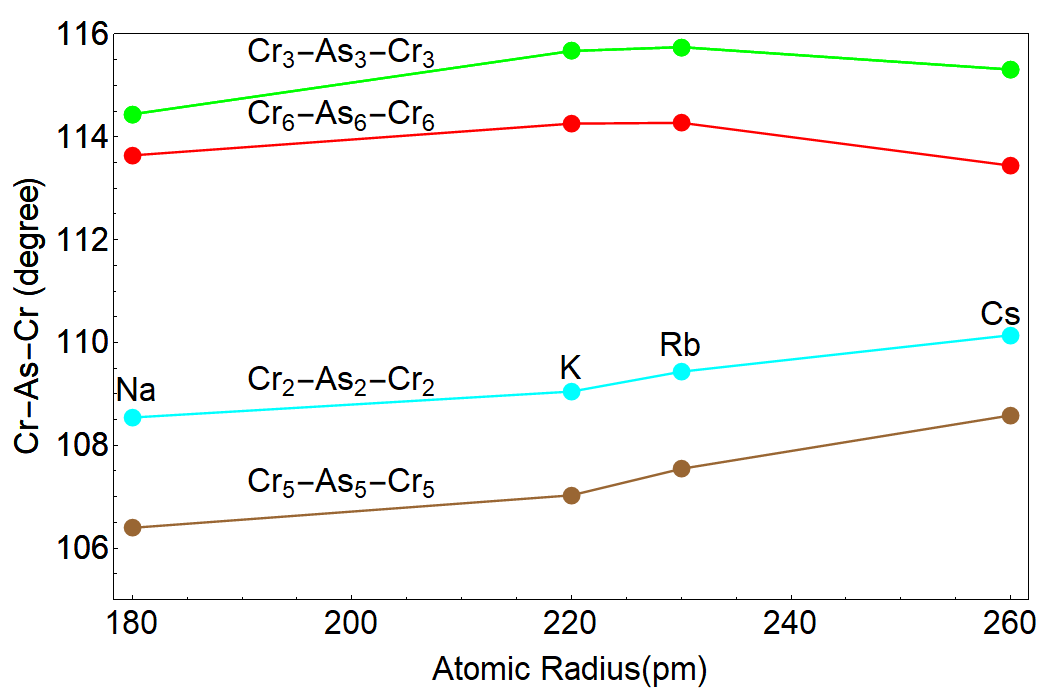}
\caption{Behavior of the four bonding angles Cr-As-Cr for different choices of the cation. The value of the Coulomb repulsion is $U=0.75\,$eV.}
\label{bondingangles}
\end{figure}

As a next step, we have evaluated the difference in energy of the NM state with respect to the FM one for different choices of the cation. We have performed the calculation for $U=0.75\,$eV, which guarantees an intrachain magnetism. In Fig.~\ref{EnvsBonding} we show that the stability of the interchain ferromagnetic coupling follows a non-monotonic trend with increasing the ion size. Actually it grows from Na to Rb, where assumes the maximum value, and then decreases for the Cs case. In order to explore the correlation between this trend and the features of the crystal structure, we have also evaluated the evolution of the bonding angles Cr-As-Cr upon alkali atom substitution.
In Fig.~\ref{bondingangles} we show that two of the four different bonding angles, that is, the Cr$_3$-As$_3$-Cr$_3$ e Cr$_6$-As$_6$-Cr$_6$ angles, display an evolution as a function of the ion substitution that follows the same trend (to make this more explicit, the behavior of the angle Cr$_6$-As$_6$-Cr$_6$ has been plotted in Fig.~\ref{EnvsBonding} together with the NM-FM energy difference). We thus deduce that the increase of these two angles may favor the interchain ferromagnetic coupling. 

We can conclude that the cation substitution acts to modify the interchain structure, thus affecting the stability of ferromagnetism among the chains; this turns out to be favored in the Rb-based compound, while it is indeed hindered in the Cs case.  We notice that recent NMR measurements suggested that Rb$_{2}$Cr$_{3}$As$_{3}$ may be indeed very close to a FM quantum critical point \cite{Yang15}, while no evidence of enhancement of FM fluctuations is actually seen in the case of Cs$_{2}$Cr$_{3}$As$_{3}$ \cite{Zhi16}. The mechanism which relates the structural deformation associated to the bonding angles to interchain ferromagnetism deserves further investigation.
\\
\\

\section{Strain to induce magnetism in K$_2$C\lowercase{r}$_3$A\lowercase{s}$_3$}

In this section we show that it is possible to tune the system from NM to interchain FM ground states by applying a compressive strain along a specific in-plane direction.
Since the magnetic properties of these compounds are very sensitive to the modification of the total volume, we will analyze the interplay between structural and magnetic degrees of freedom in the case of fixed volume, as already explained in Section II.\\

We choose to focus on the reference case of K$_2$Cr$_3$As$_3$ and explicitly compare it with KCr$_3$As$_3$ sharing with K$_2$Cr$_3$As$_3$ quasi-one-dimensional structural features. However, in the latter no superconductivity was found while a spin-glass-like transition at $T_N = 5\,$K was revealed by magnetic susceptibility measurements  \cite{Bao15b}. This value is very close to the superconducting critical temperature of K$_2$Cr$_3$As$_3$, indicating that the energy scales involved in the formation of the two phases are indeed comparable and  that magnetism may be detrimental to the development of the superconducting phase. 
We also mention that KCr$_3$As$_3$ also presents a superconducting phase \cite{Mu17,Liu18}, though recently Taddei {\it et al.} \cite{Taddei19} have shown that the emerging of this phase is due to the charge doping via H intercalation.

\begin{figure}[]
\centering
\includegraphics[width=7 cm, angle=0]{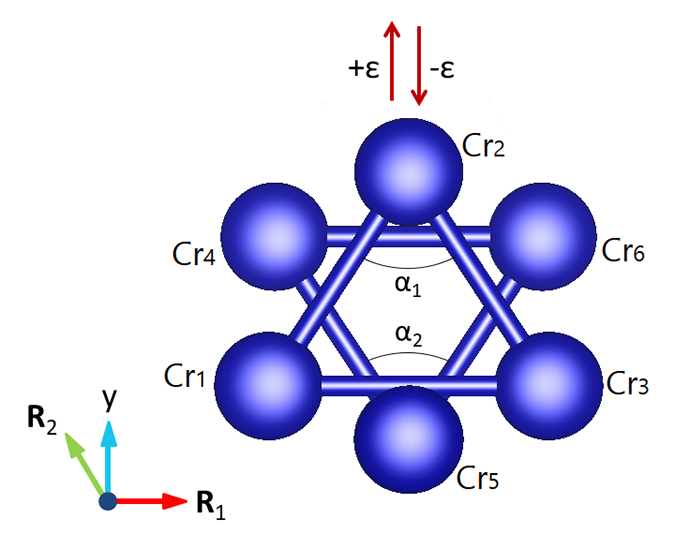}
\caption{Graphic representation of the strain $\varepsilon$ applied orthogonally to the basis of the isosceles triangles in the chain. A negative value of $\varepsilon$ indicates a compressive strain of the y-component of the \textbf{R}$_2$ vector, while a positive one indicates a tensile strain of the y-component of the \textbf{R}$_2$ vector.}
\label{strain1}
\end{figure}

We start by referring to the lattice vectors of the undistorted unit cell \textbf{R}$_1$=(a,0,0), \textbf{R}$_2$=(-a/2,$\sqrt{3}a/2$,0), and \textbf{R}$_3$=(0,0,c). The strain field $\varepsilon$ applied to the unit cell is defined as the continuous deformation of the in-plane lattice vectors \textbf{R}$_1$ and \textbf{R}$_2$. 
A negative value of $\varepsilon$ corresponds to a compressive strain of the $y$-component of the \textbf{R}$_2$ vector, while a positive value corresponds to a tensile one.
Since the volume is fixed to the value $\frac{\sqrt{3}ca^2}{2}$, the lattice vectors then become
\begin{eqnarray*}
\textbf{R}_1 & = & \left(\frac{a}{1+\varepsilon},0,0 \right)\\
\textbf{R}_2 & = & \left(-\frac{a}{2},\frac{\sqrt{3}a(1+\varepsilon)}{2},0 \right)\\
\textbf{R}_3 & = & \left(0,0,c \right)
\end{eqnarray*}

where $\epsilon$ is adimensional and defined as $\varepsilon=\Delta a/a$. The effect of this compressive (tensile) strain of the y-component of the \textbf{R}$_2$ vector enhances (decreases) the apex $\alpha$ angles as shown in Fig.~\ref{strain1}.
Due to the strong interplay between structural and magnetic degrees of freedom in this class of materials, this kind of strain is expected to have significative effects on the magnetic properties.
The modules of the lattice vectors of K$_{2}$Cr$_{3}$As$_{3}$ for different values of the strain $\varepsilon$ are reported in Table \ref{Tab2}. 

\begin{figure}[]
\centering
\includegraphics[width=\columnwidth, angle=0]{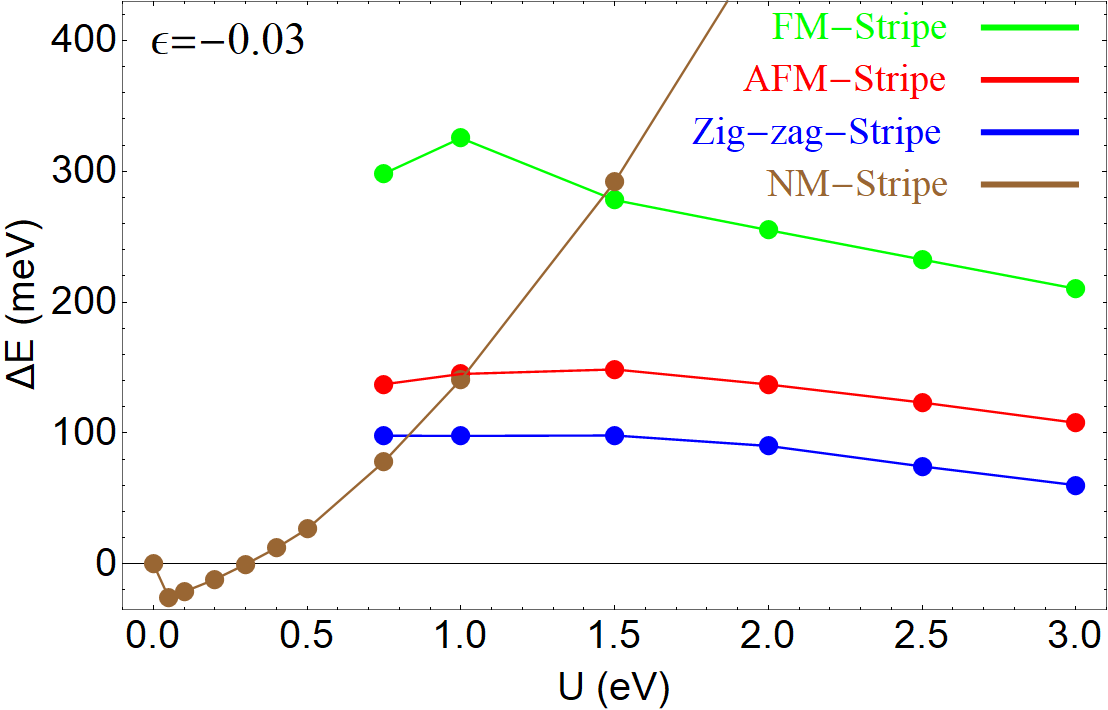}
\caption{Energies per Cr-atom of the FM, AFM, zig-zag and NM states measured with respect to the stripe state energy, plotted as functions of the Coulomb repulsion for the compound K$_2$Cr$_3$As$_3$. A compressive strain $\varepsilon=-$0.03 is applied to the system. At low values of $U$ some data are missing due to the lack of convergence of the numerical procedure in that regime.}
\label{EC_K2strain3}
\end{figure}

The comparative analysis of the most favorable magnetic configurations within the chain for negative/positive strain demonstrates that the ground state is still NM below a critical $U_c$, and that above this threshold value a stripe phase is favored (see Fig.~\ref{EC_K2strain3}). Compared with the unstrained case, this value is lowered to $\sim 0.3\,$eV when a 3$\%$ negative strain is considered, and a further slight reduction is observed when $\varepsilon$ is further increased.

\begin{figure}[]
\centering
\includegraphics[width=\columnwidth, angle=0]{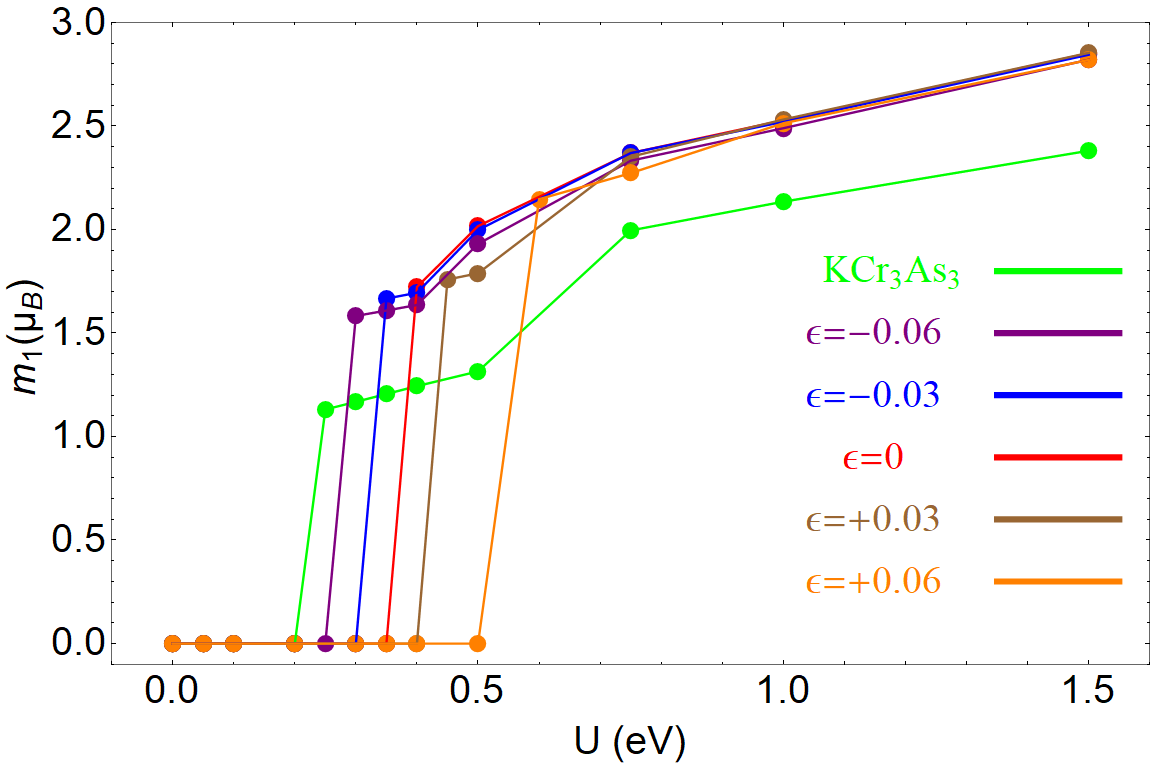}
\caption{Comparison of the magnetic moments of the Cr$_1$ atom in the ground state for different choices of compressive/tensile strain applied to K$_{2}$Cr$_{3}$As$_{3}$.}
\label{Ground_MagMom_Strain}
\end{figure}

\begin{figure}[]
\centering
\includegraphics[width=\columnwidth, angle=0]{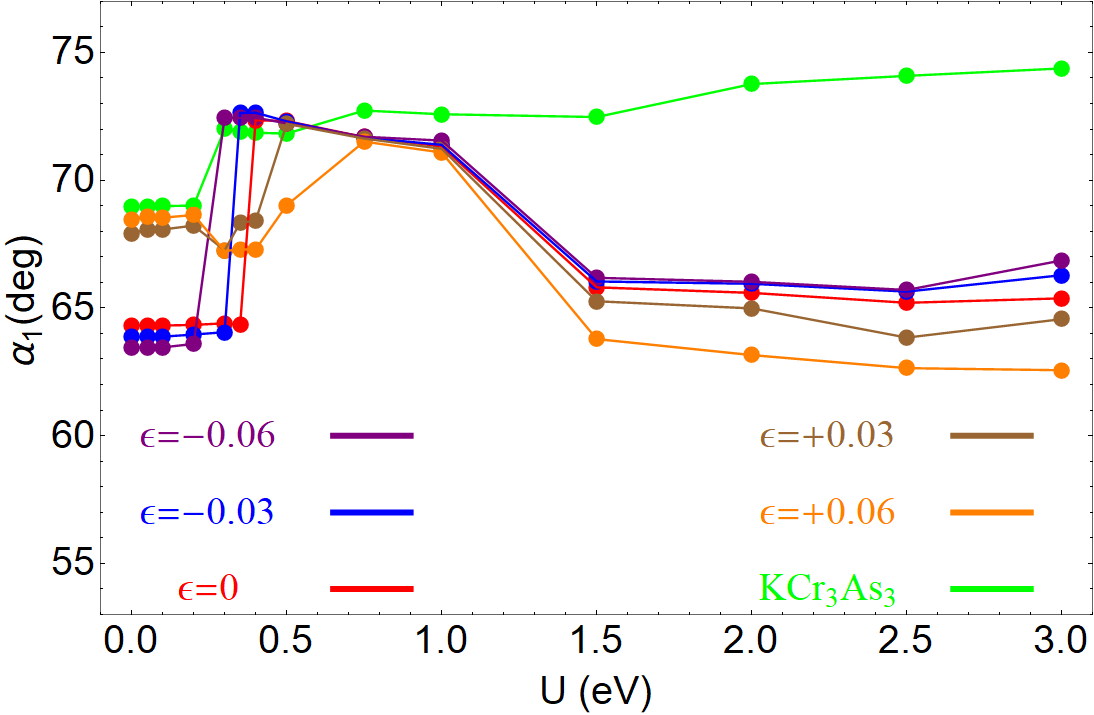}
\caption{Comparison of the $\alpha_1$ angle in the ground state for different choices of compressive/tensile strain applied to K$_{2}$Cr$_{3}$As$_{3}$.}
\label{angles_comparisonstrain}
\end{figure}

In order to provide a systematic comparison of the modifications to the ground-state magnetic and structural properties induced by a compressive/tensile strain, we report in Figs.~\ref{Ground_MagMom_Strain} and \ref{angles_comparisonstrain} the evolution with $U$ of the magnetic moment at the representative Cr$_1$ site and of the angle $\alpha_1$ in the ground state, together with the analogous results recently obtained for KCr$_{3}$As$_{3}$ \cite{Cuono20}.

From the inspection of Fig.~\ref{Ground_MagMom_Strain}, we notice that the compressive and the tensile strain both allow to tune the transition from a non-magnetic phase with vanishing moment to a  magnetic one with $m_1\ne 0$ (turning out to be the stripe phase), but following an opposite trend. Tensile strain raises the value of $U$ at which a finite moment appears, whereas a compressive strain tends to reduce it. In particular, for $\varepsilon$=$-$0.06 this value is close to the critical one marking the transition to the stripe state in KCr$_{3}$As$_{3}$. In this context, it is significant that the angle $\alpha_1$ shows an increase at the transition, particularly sharp for compressive strain, as one can see from Fig.~\ref{angles_comparisonstrain}.

Next we turn to the analysis of the lowest energy configurations when the coupling between magnetic chains is considered. The collinear configurations is assumed within the chains and the calculations are performed within the PBEsol approximation. The energies of the NM, FM and AFM configurations for K$_{2}$Cr$_{3}$As$_{3}$ are reported in Tab.~\ref{Tabamongchainsstrain} for different values of the strain. We have fixed for the Coulomb repulsion the value $U=0.3\,$eV, for which KCr$_{3}$As$_{3}$ is predicted to be magnetic, while K$_2$Cr$_{3}$As$_{3}$ is not \cite{Cuono20}.
The analysis of the interchain interactions in K$_{2}$Cr$_{3}$As$_{3}$ confirms the trend emerging from the study of magnetism inside the single chain. Namely, a compressive strain leads the system towards a magnetic phase, while a tensile strain stabilizes the non-magnetic ground state. A value $\varepsilon$=$-$0.06 favors K$_{2}$Cr$_{3}$As$_{3}$ to be magnetic at $U=0.3\,$eV. The ground state is the interchain FM configuration, with the macrospins of the chains oriented in the same direction.

Hence, our results seem to indicate that the instability towards a ferrimagnetic phase can be tuned by the application of a compressive strain. This could provide important insights on the interplay between superconductivity and magnetism, possibly providing support to the idea that the superconducting phase arises from the suppression of the magnetic order, as controlled by the strain as a tunable parameter.

\noindent 
\begin{table}[t!]
\begin{centering}
\begin{tabular}{|c|c|c|c|c|}
\hline 
  K$_{2}$ with strain  $\varepsilon$  & $\varepsilon$= $-$ 0.06 &  $\varepsilon$= $-$ 0.03 & $\epsilon$=+0.03 & $\epsilon$=+0.06 \tabularnewline
\hline 
$|\textbf{R}_1|\;\,$(\AA)  & 10.6204  &  10.2920 & 9.6924 & 9.4181 \tabularnewline
\hline 
$|\textbf{R}_{2}|\;\,$(\AA)  & 9.5375  &  9.7594 & 10.2086 & 10.4357 \tabularnewline
\hline  
$|\textbf{R}_3|\;\,$(\AA)  & 4.2304  & 4.2304 & 4.2304 & 4.2304 \tabularnewline
\hline
\end{tabular}
\par\end{centering}
\caption{Modules of the lattice vectors of K$_{2}$Cr$_{3}$As$_{3}$ when the strain is applied.}
\label{Tab2} 
\end{table}

\noindent 
\begin{table}[t!]
\begin{centering}
\begin{tabular}{|c|c|c|c|c|c|}
\hline 
& $\epsilon$=$-$0.06 & $\epsilon$=$-$0.03 & $\epsilon$=0 & $\epsilon$=+0.03 & $\epsilon$=+0.06 \tabularnewline
\hline  
NM  & 5.33  & 0 & 0 & 0   & 0  \tabularnewline
\hline 
FM  & 0  & 0.25 & 6.92 & 15.25  & 26.75  \tabularnewline
\hline 
AFM  & 3.92  & 4.08 & 10.83 & 19.75  & 32.00  \tabularnewline
\hline  
\end{tabular}
\par\end{centering}
\caption{Energy differences (meV) of K$_{2}$Cr$_{3}$As$_{3}$ for NM, FM and AFM coupling between the chains and different values of the strain. Inside the chain the stripe configuration is assumed to be the ground and the value of the Coulomb interaction is fixed at $U=0.3\,$eV.}
\label{Tabamongchainsstrain} 
\end{table}

\section{Conclusions}
We analyzed the ground state of the series A$_{2}$Cr$_{3}$As$_{3}$  (A=Na, K, Rb, Cs) and predicted a collinear stripe configuration within the DWSN, which allows to attribute a
net magnetic moment to each chain. Due to interchain ferromagnetic coupling, all the compounds are close to a ferrimagnetic phase in the region of moderate values of the Coulomb repulsion $U$. The occurrence of such collinear magnetic state has an important interplay with the distortion of the triangles which reduces the frustration of the antiferromagnetic couplings between the nearest neighbor Cr atoms.

Such behavior has been proved to be robust against the variation of the chemical pressure induced by the change of the cation among the different members of the family. Notably, we have shown that the strength of the interchain ferromagnetic coupling has a non-monotonic behavior as a function of the atomic radius of the alkali metals. In particular, the stability of the interchain ferromagnetic coupling is gradually increased when changing from Na to Rb, while it is reduced for the Cs compound, in agreement with recent experimental observations \cite{Yang15,Zhi16}. We relate this behavior to the Cr-As-Cr bonding angles along the chain, this being a key factor controlling the tendency towards the interchain ferromagnetism.

As far as strain is concerned, we demonstrated that uniaxial compressive strain applied orthogonally to the basis of the isosceles triangles tends to increase the apex angles. Confirming the strong interplay between structural properties and magnetism in the above-mentioned compounds, strain has been shown to significantly affect the transition from the NM to the ferrimagnetic phase in the regime of moderate electron correlations. 

In conclusion, our results clearly show that the compounds of the family A$_{2}$Cr$_{3}$As$_{3}$ (A= Na, K, Rb, Cs) are close to an interchain FM instability, which is significantly affected by structural effects. Our analysis can thus prove to be relevant in the study of the interplay between magnetism and superconductivity experimentally detected in this class of compounds \cite{Wang16,Xu20}.

\section{Acknowledgments}

The authors acknowledge A. Galluzzi and M. Polichetti for useful discussions.
The work is supported by the Foundation for Polish Science through the International Research Agendas program co-financed by the European Union within the Smart Growth Operational Programme. 
G.C. acknowledges financial support from "Fondazione Angelo Della Riccia". 
X. M. was sponsored by the National Natural Science Foundation of China (No. 11864008).
We acknowledge the access to the computing facilities of the Interdisciplinary Center of Modeling at the University of Warsaw, Grant No. G73-23 and G75-10. We acknowledge the CINECA award under the ISCRA initiatives IsC69 "MAINTOP", IsC76
"MEPBI" and IsC81 "DISTANCE" Grant for the availability of high-performance computing resources and support.

\section{Appendix A: \\  Results for N\lowercase{a}$_{2}$C\lowercase{r}$_{3}$A\lowercase{s}$_{3}$ and R\lowercase{b}$_{2}$C\lowercase{r}$_{3}$A\lowercase{s}$_{3}$}

Here we report the results concerning the energies, the angle $\alpha_{1}$ and the magnetic moment $m_{1}$ for the other compounds of the family A$_{2}$Cr$_{3}$As$_{3}$. 

\begin{figure}[]
\centering
\includegraphics[width=\columnwidth, angle=0]{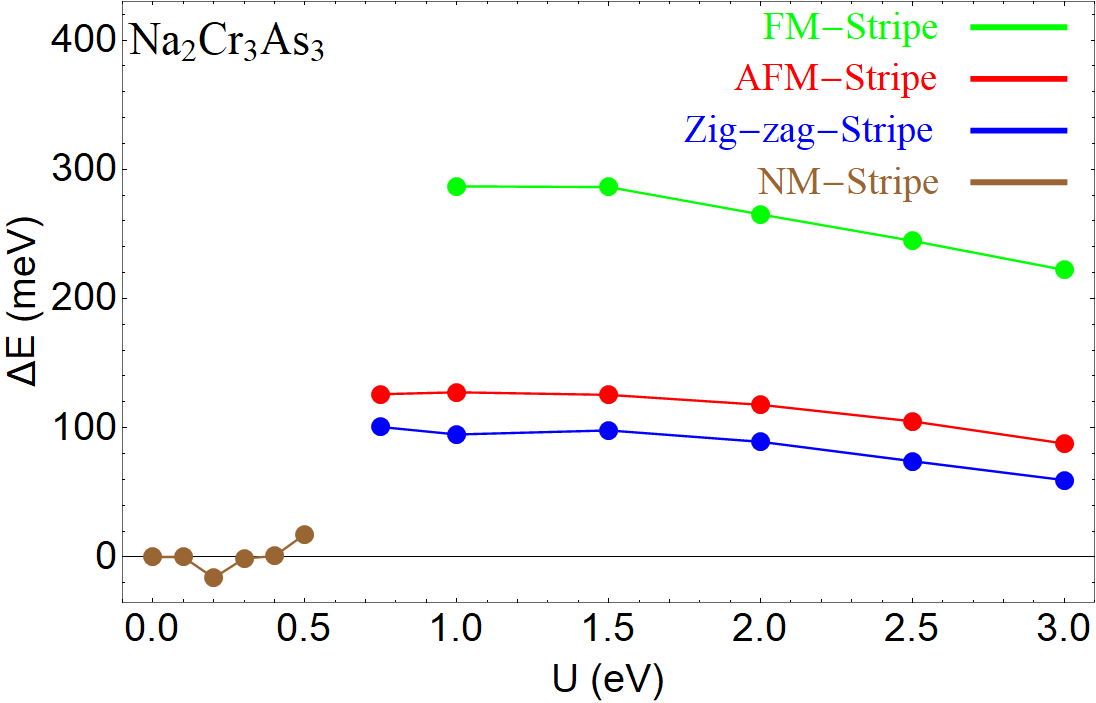}
\caption{Energies per Cr-atom of the FM, AFM, zig-zag and NM states measured with respect to the stripe state energy, plotted as functions of the Coulomb repulsion for the compound Na$_2$Cr$_3$As$_3$.
At low values of $U$ some data are missing due to the lack of convergence of the numerical procedure in that regime.}
\label{EC_Na}
\end{figure}

\begin{figure}[]
	\centering
	\includegraphics[width=\columnwidth, angle=0]{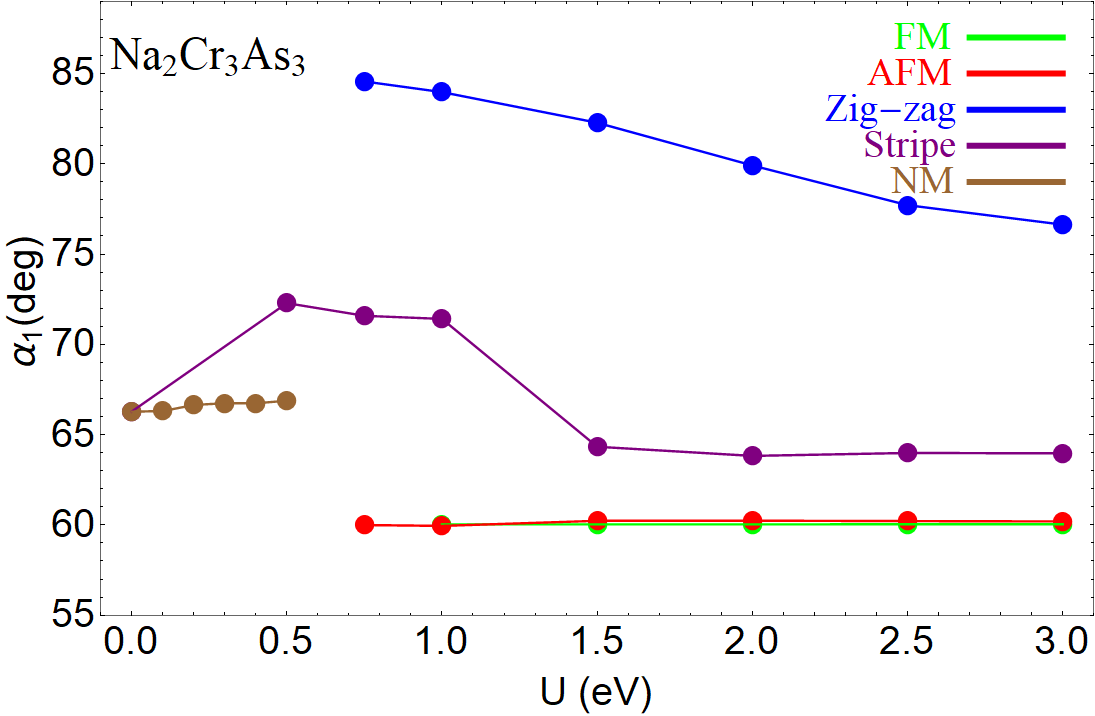}
	\caption{ Angle $\alpha_{1}$ as a function of the Coulomb repulsion for the compound Na$_2$Cr$_3$As$_3$.
	At low values of $U$ some data are missing due to the lack of convergence of the numerical procedure in that regime.}
	\label{angles_Na}
\end{figure}

\begin{figure}[]
\centering
\includegraphics[width=\columnwidth,  angle=0]{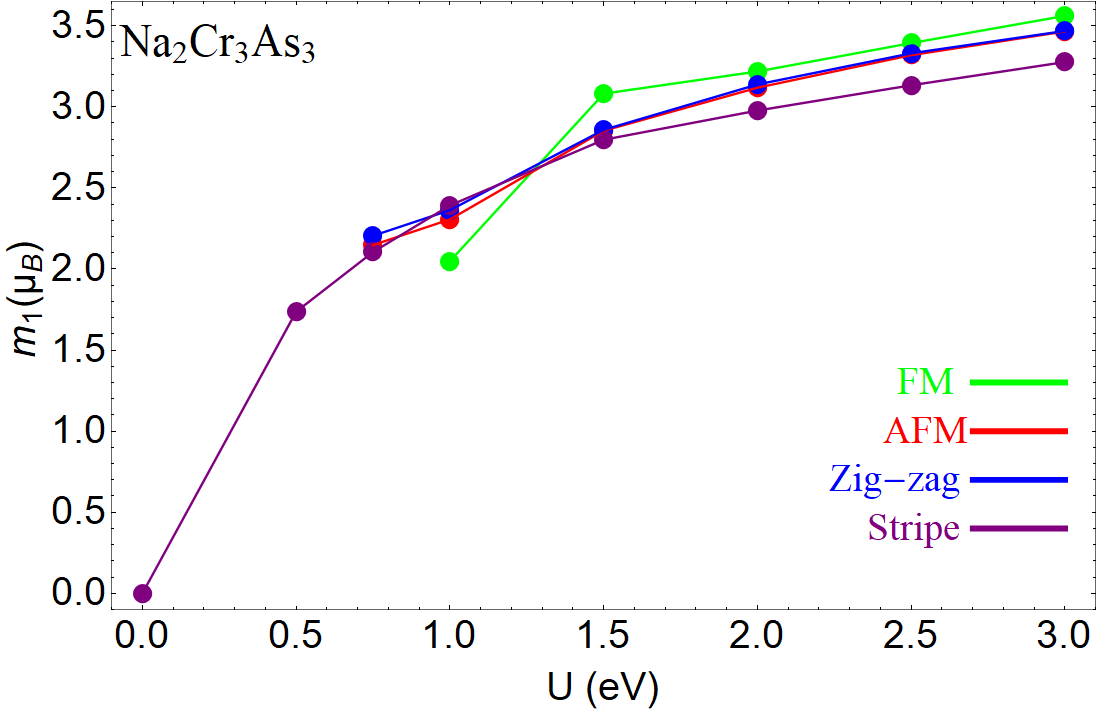}
\caption{Magnetic moment of the Cr$_{1}$ ion, as a function of the Coulomb repulsion for the compound Na$_2$Cr$_3$As$_3$. 
At low values of $U$ some data are missing due to the lack of convergence of the numerical procedure in that regime.}
\label{magneticmoment_Na}
\end{figure}

\begin{figure}[]
\centering
\includegraphics[width=\columnwidth, angle=0]{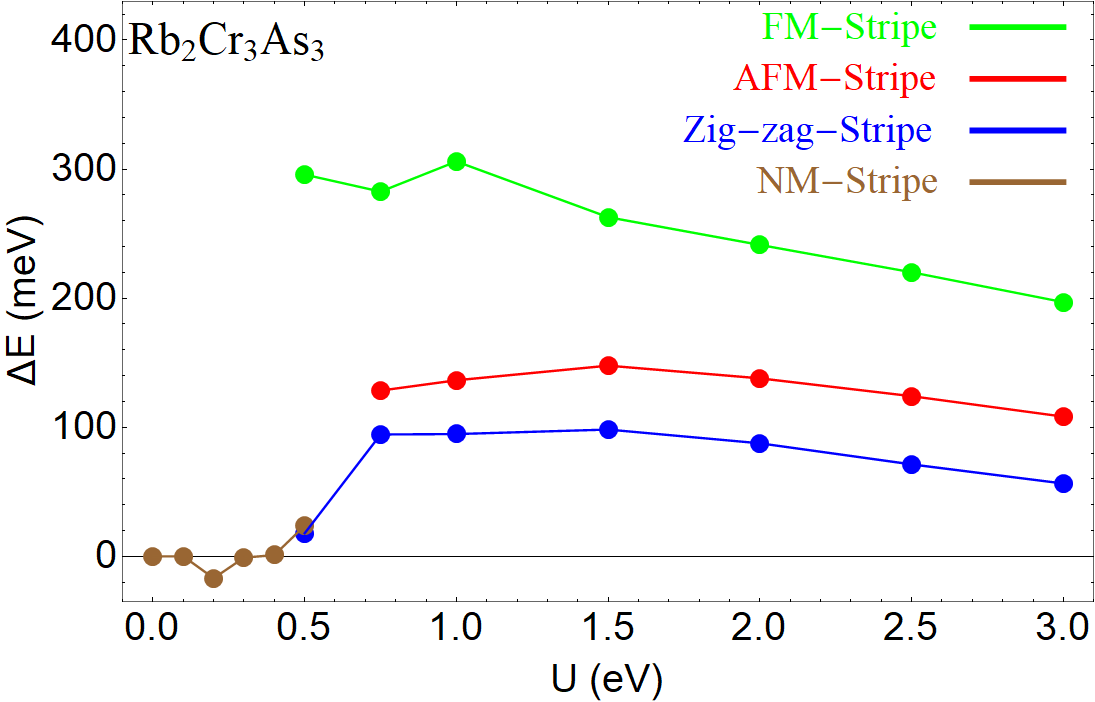}
\caption{ Same as in Fig.~\ref{EC_Na} for the compound Rb$_{2}$Cr$_{3}$As$_{3}$.}
\label{EC_Rb}
\end{figure}

The energies of the various configurations investigated, measured with respect to the one of the stripe state, are plotted as functions of the Coulomb repulsion $U$ in Figs.~\ref{EC_Na} and \ref{EC_Rb} for Na$_{2}$Cr$_{3}$As$_{3}$ and Rb$_{2}$Cr$_{3}$As$_{3}$, respectively. 
The case of the K$_{2}$Cr$_{3}$As$_{3}$ is instead reported in Ref.~\cite{Cuono20}.

\begin{figure}[]
	\centering
	\includegraphics[width=\columnwidth, angle=0]{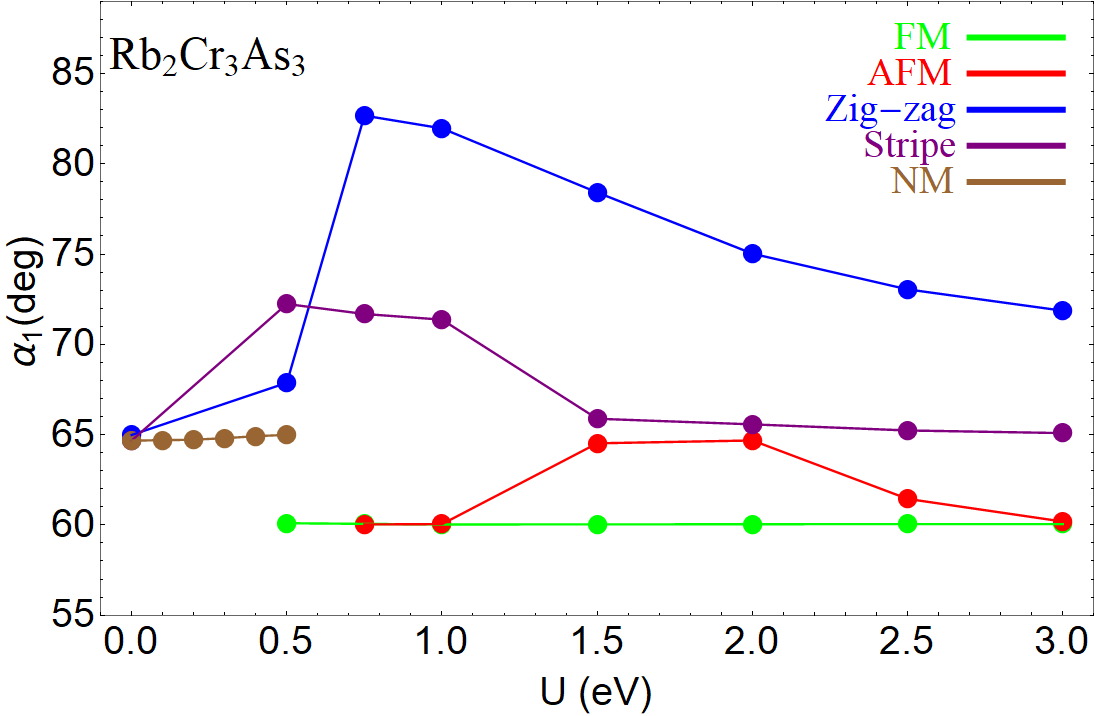}
	\caption{ Same as in Fig.~\ref{angles_Na} for the compound Rb$_{2}$Cr$_{3}$As$_{3}$.
	}
	\label{angles_Rb}
\end{figure}

\begin{figure}[]
\centering
\includegraphics[width=\columnwidth,  angle=0]{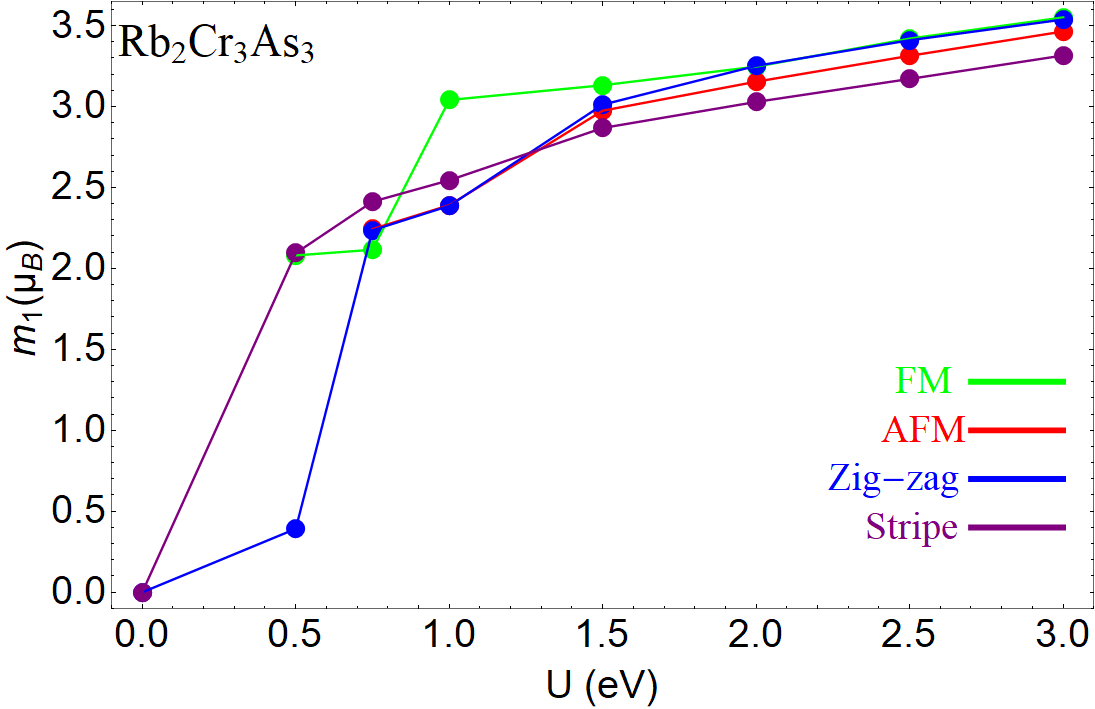}
\caption{Same as in Fig.~\ref{magneticmoment_Na} for the compound Rb$_{2}$Cr$_{3}$As$_{3}$.}
\label{magneticmoment_Rb}
\end{figure}

Our results show a similar behavior for the two compounds, this indicating that the chemical pressure due to the change of the cation alters only slightly the nature of the magnetic configuration within the chain. For all the compounds the ground state is non-magnetic for values of the Coulomb repulsion $U<U_c=0.4\,$eV, then becoming the stripe one for $U\geq U_c$.
On the other hand the energy in the ferromagnetic configuration is always larger than in the other phases, this situation remaining the same for all the compounds.

The angle $\alpha_1$ as a function of $U$ is plotted in Figs.~\ref{angles_Na} and \ref{angles_Rb} for Na and Rb cations,  respectively. The various configurations always correspond to distorted triangles, except for the fully FM one, and, in the case of Na$_{2}$Cr$_{3}$As$_{3}$, also for the AFM one.
For all the compounds, the increase of $U$ above a value approximately equal to 1.5 eV does not produce significant variations of $\alpha_1$, the only exception being the zig-zag configuration for which $\alpha_1$ decreases appreciably as $U$ is increased. 
The magnetic moment of the Cr$_1$ ion is reported for the two cases A=Na, Rb in Figs.~\ref{magneticmoment_Na} and \ref{magneticmoment_Rb}, respectively.
For $U>1.5\,$eV it approaches the maximum predicted value for Cr ions in A$_{2}$Cr$_{3}$As$_{3}$, that is 10/3 $\mu_B$. 
On the other hand, for small values of the Coulomb repulsion the magnetic moment tends to vanish, in agreement with the fact that the ground state in this regime is non-magnetic.

\section{Appendix B: Magnetic exchanges}

In this appendix, we map the DFT results into the Heisenberg model, providing an estimation of the exchange couplings and explaining the limitations of the method when applied to the class of materials under consideration. First we provide the results for the magnetic exchanges in K$_2$Cr$_3$As$_3$ as functions of $U$, then we plot their evolution as a function of the strain, as described in the main text.

The mapping of the DFT results on the Heisenberg model can be obtained calculating the energy of several magnetic configurations. 
The reliability of this method depends on the values of the magnetic moments that should be constant in each of the magnetic configurations. While this holds
for $U\gtrsim\,1.5\,$eV, we have found that the magnetic moment appreciably varies for $U<1.5\,$eV. However, our results show that it is still possible to perform the mapping of the DFT results into the Heisenberg model for 0.75\,eV $\lesssim U<\,$1.5\,eV, but for this class of materials we are on the verge of the applicability of the mapping to the Heisenberg model. Since we know that the magnetic coupling should go to zero for U$<$0.3 eV, we extrapolate the results from $U=0.75\,$eV to $U=0.3\,$eV to have an indication of the magnetic coupling in the range of $U$ that better describes the properties of the system.

\begin{figure}
  \includegraphics[scale=0.45]{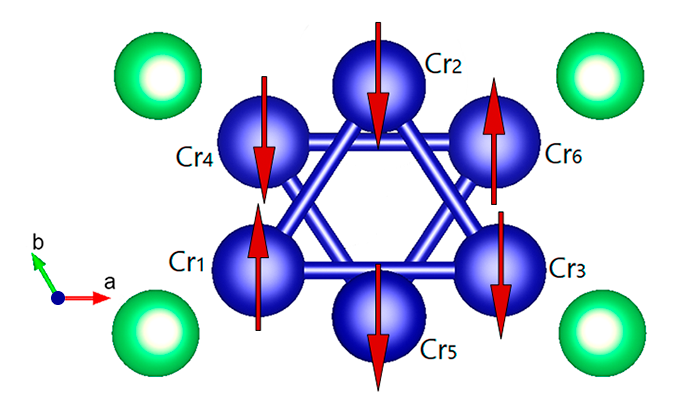}
	\caption{Arrangements of the Cr-spin of the magnetic configuration investigated in addition to the previous ones discussed in the main text in order to obtain the magnetic exchanges. The new magnetic configuration has the Cr atoms at the basis with opposite spin direction. $J_a$ is the magnetic coupling between the Cr-atoms at the basis and those at the apex of the isosceles triangles, as for example Cr$_1$ and Cr$_2$; $J_b$ is the magnetic coupling between the two Cr-atoms at the basis, as Cr$_1$ and Cr$_3$. $J_c$ is the inter-plane coupling constant, relating for example to atoms Cr$_1$ and Cr$_4$.
	}
	\label{newconfiguration}
\end{figure}

To perform the mapping and have an estimation of the nearest-neighbor magnetic coupling, we assume that the two interlayer magnetic couplings are equal and we calculate the energy for a new magnetic configuration in addition to the previous ones shown in Fig. \ref{configurations} and investigated in the main text.
The new magnetic configuration has the Cr atoms at the basis with opposite spin direction as shown in Fig. \ref{newconfiguration}.
We exclude the ferromagnetic phase configuration for the mapping because its equation is linearly dependent on the others in the
linear equation system.

As in Ref. \onlinecite{Cuono20}, we indicate with $J_a$ the magnetic coupling between the Cr-atoms at the basis and those at the apex of the isosceles triangles, as for example Cr$_1$ and Cr$_2$ of Fig. \ref{newconfiguration}, while $J_b$ is the magnetic coupling between the two Cr-atoms at the basis, as Cr$_1$ and Cr$_3$. $J_c$ is the inter-plane coupling constant, relating for example to atoms Cr$_1$ and Cr$_4$.

\begin{figure}
  \includegraphics[width=\columnwidth,  angle=0]{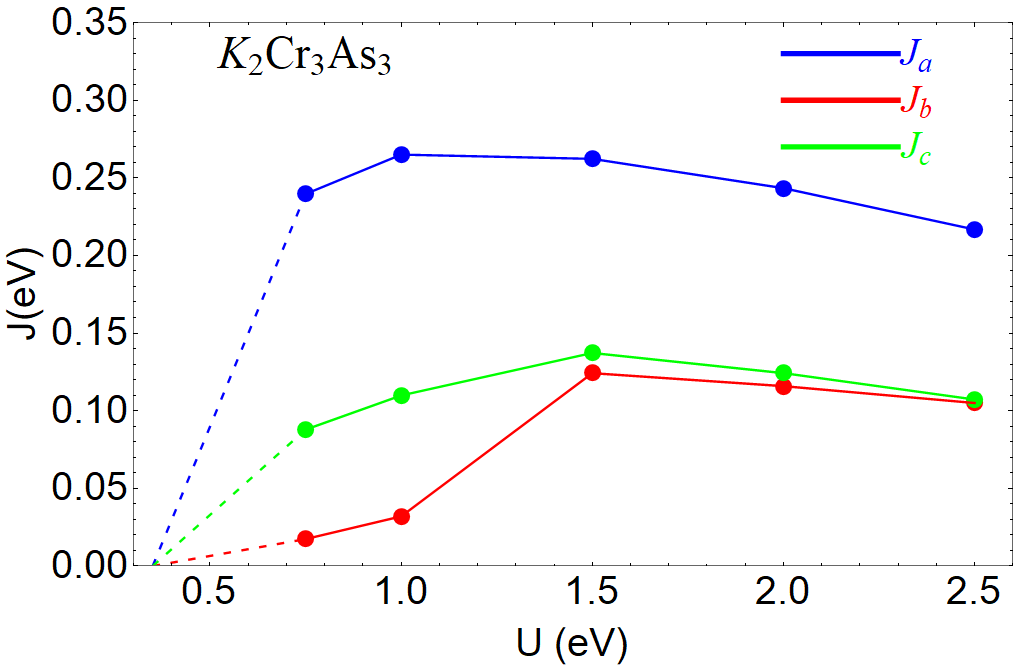}
	\caption{Magnetic exchanges for K$_2$Cr$_3$As$_3$ as functions of the Coulomb repulsion $U$. 
	$J_a$, $J_b$ and $J_c$ are three magnetic exchanges defined in the text.
	The dashed line indicates the extrapolation in the region where the mapping is not possible. 
	}
	\label{Jvalues_U}
\end{figure}

The magnetic exchanges for K$_2$Cr$_3$As$_3$ as functions of $U$ are reported in Fig.\ref{Jvalues_U}.
All the nearest-neighbor magnetic exchanges are antiferromagnetic for every value of $U$, including the value $U=2\,$eV that is also reported in the literature for the undistorted case \cite{Wu15}.
When the coupling is antiferromagnetic, the value of $J$ is positive in our convention.
The in-plane magnetic coupling $J_a$ is larger than $J_b$, this stabilizing the collinear magnetic configuration, as reported in Ref. \cite{Cuono20}. Moreover, the interlayer magnetic coupling is larger than the previously calculated magnetic couplings  \cite{Wu15}, this being presumably due to the increase of the distortions that reduces the strength of the interlayer Cr-Cr bonds.

\begin{figure}{b}
  \includegraphics[width=\columnwidth,  angle=0]{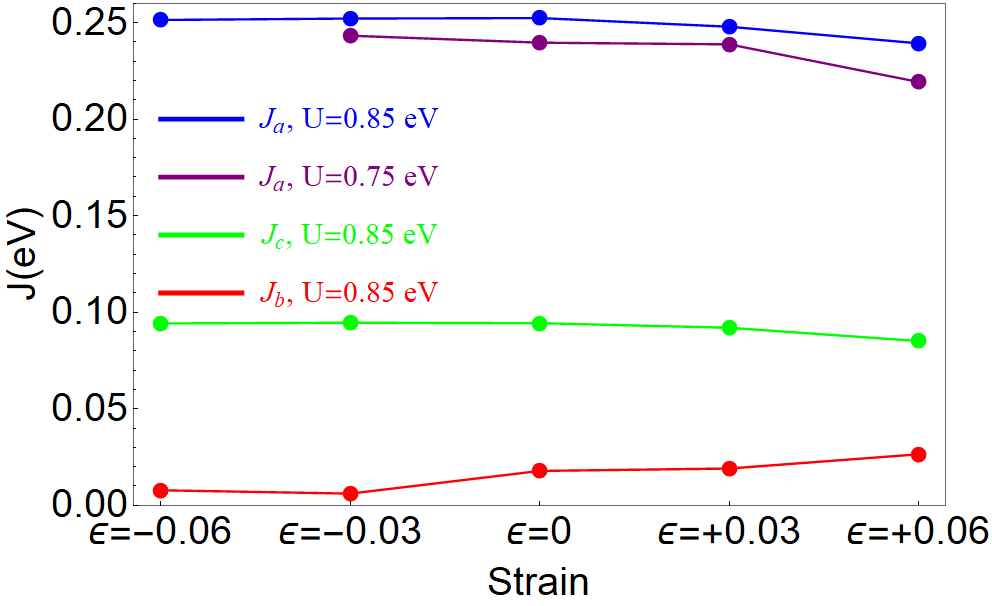}
	\caption{Magnetic exchanges of K$_2$Cr$_3$As$_3$ as functions of the strain for $U$=0.85 eV and $J_a$ in the case of $U$=0.75 eV. 
	$J_a$, $J_b$ and $J_c$ are three magnetic exchanges defined in the text.}
	\label{Jvalues_strain}
\end{figure}

In Fig. \ref{Jvalues_strain}, we plot the magnetic exchanges as functions of the strain for $U=0.85\,$eV, this being the lowest value of $U$ where the mapping is possible for every value of the strain. While the magnetic exchanges are very sensitive to the Coulomb repulsion $U$, the dependence on the strain is much weaker, as expected from the other results discussed in the main text.
We also report the behavior of $J_a$ for $U=0.75\,$eV showing that the main magnetic coupling $J_a$ increases as a function of $U$ for every value of the strain. 
We see that a tensile (positive) strain brings towards a non-magnetic phase, as stated in the main text of the paper, see Fig. \ref{Jvalues_strain}.
We cannot map the $J$'s for $U=0.3\,$eV, but we expect that for negative value of the strain the magnetic couplings are still sizeable, in such a way that we can observe a long-range magnetic order in the presence of interchain magnetic coupling \cite{Cuono20}.

As far as the estimation of the critical temperature is concerned, since the system is quasi-one-dimensional with frustrated magnetism, a mean-field approach fails to provide a correct estimation of $T_c$. Rather, we should include in the calculation the effect of the inter-chain exchange interactions, but this is beyond the scope of the present paper. 
\\

\end{document}